\documentclass[preprint]{aastex}
\usepackage{graphics} 

\shorttitle{AL in SDSS1125}
\shortauthors{Shi \textit{et al.}}

\begin{document}

\title{The Redshifted Hydrogen Balmer and Metastable \ion{He}{1} Absorption Line System in Mini-FeLoBAL Quasar SDSS J112526.12+002901.3: A Parsec Scale Accretion Inflow?}

\author{Xi-Heng Shi\altaffilmark{1,2}, Peng Jiang\altaffilmark{1}, Hui-Yuan Wang\altaffilmark{2}, Shao-Hua Zhang\altaffilmark{1}, Tuo Ji\altaffilmark{1}, Wen-Juan Liu\altaffilmark{3}, and Hong-Yan Zhou\altaffilmark{2,1}}

\altaffiltext{1}{Polar Research Institute of China, Jinqiao Road 451, Shanghai 200136, China}
\altaffiltext{2}{Key Laboratory for Research in Galaxies and Cosmology, Department of Astronomy, University of Science and Technology of China, Chinese Academy of Sciences, Hefei, Anhui 230026, China}
\altaffiltext{3}{Yunnan Observatories, Chinese Academy of Sciences, Kunming, Yunnan 650011, China}

\begin{abstract}

The accretion of interstellar medium onto the central super massive black holes is widely accepted as the source of the gigantic energy released by the active galactic nuclei. But few pieces of observational evidence have been confirmed directly demonstrating the existence of the inflows. The absorption line system in the spectra of quasar SDSS J112526.12+002901.3 presents an interesting example, in which the rarely detected hydrogen Balmer and metastable \ion{He}{1} absorption lines are found redshifted to the quasar's rest frame along with the low-ionization metal absorption lines \ion{Mg}{2}, \ion{Fe}{2}, etc. The repeated SDSS spectroscopic observations suggest a transverse velocity smaller than the radial velocity. The motion of the absorbing medium is thus dominated by infall. The \ion{He}{1}* lines present a powerful probe to the strength of ionizing flux, while the Balmer lines imply a dense environment. With the help of photoionization simulations, we find the absorbing medium is exposed to the radiation with ionization parameter $U\approx 10^{-1.8}$, and the density is $n(\mathrm{H})\approx 10^9\ \mathrm{cm}^{-3}$. Thus the absorbing medium is located $\sim 4\ \mathrm{pc}$ away from the central engine. According to the similarity in the distance and physical conditions between the absorbing medium and the torus, we strongly propose the absorption line system as a candidate for the accretion inflow which originates from the inner surface of the torus.

\end{abstract}


\section{Introduction}

Active Galactic Nucleus (AGN) is one of the most luminous objects in the universe. Observed as bright stellar-like point source, it is located in the core of massive galaxy. The extremely high luminosity and rapid variability require that gigantic energy is generated and released in a quite small volume with linear size not larger than a few parsecs. The accretion of interstellar medium onto super-massive black holes (SMBHs) is widely accepted as the process driving these central engines (Lynden-Bell 1969). The gravitational potential of the infalling matter is transfered into radiation by the viscous stress on the accretion disks surrounding the SMBHs (Rees 1984). Moreover, massive outflows from the inner part of the accretion disks are expected, not only to carry away the angular momentum of infalling matter to maintain the accretion, but also to explain the observed correlation between the properties of central SMBHs and their host galaxies. This correlation requires a mechanism of feedback from the central engines to regulate the star formation in the host galaxies (Granato \textit{et al.} 2004; Scannapieco \& Oh 2004; Hopkins \textit{et al.} 2008).

The feedback is believed to have solid direct observational basis, since both the asymmetric emission line profiles (Gaskell 1982; Richards \textit{et al.} 2002; Wang \textit{et al.} 2011) and the intrinsic absorption lines blueshifted to the quasars' rest frames (Weymann \textit{et al.} 1981) are detected and explained as the gaseous outflows. The latter, the blueshifted absorption troughs, are found with large outward velocity varying from a few thousand km s$^{-1}$ to about 0.2 $c$. Usually the profiles of these absorption troughs also display great widths, larger than $2000\ \mathrm{km\ s}^{-1}$. Thus, they are classified as broad absorption lines (BALs). A great deal of research works, for samples (Weymann \textit{et al.} 1991; Hewett \& Foltz 2003; Reichard \textit{et al.} 2003; Zhang \textit{et al.} 2010) or for individual objects (Wampler \textit{et al.} 1995; de Kool \textit{et al.} 2001, 2002; Leighly \textit{et al.} 2011; Zhang \textit{et al.} 2015a), have been published to investigate the geometry and evolution of the global structure of the outflows, or the physical conditions of individual outflowing clouds.

But the observational evidences of the more basic process, the accretion, remain questionable. Till now, few have been confirmed to directly demonstrate the existence of the accretion inflow in the vicinity of AGNs. Since the inflows are suggested originate from the torus (Vollmer \textit{et al.} 2004; Beckert \& Duschl 2004), the most possible place to discover it is the gap between the inner surface of the torus and the outer region of the accretion disk. But the ionizing flux here from the central engine is relatively weak so that the emission of the illuminated inflow would be overwhelmed by the broad emission lines. The absorption lines provide a more accessible way for the detection if the medium intercepts our line of sight (LOS) toward the central radiation source. The problem is that these inflows are expected to lie not far from the equatorial plane so that the LOS would be obscured by the torus.

Recent models tend to prefer a clumpy torus to explain the lack of $10\ \micron$ Si emission features in the spectral energy distribution (SED) of type 1 AGN (Nenkova \textit{et al.} 2002). The impact of such models to our concern is that we could thus have chances to look through the low-density part among the dense clumps especially when the angle of the LOS from the equator is relative large where the torus medium would be a bit more diffuse. The probability must be quite low. But once this occurs, the LOS could also intercepts the assumed inflow to unveil its existence by the redshifed absorption profile as a result of the inward motion.

The Sloan Digital Sky Survey (SDSS) provides the largest quasar sample, containing 105783 objects in the 7th data release for SDSS-I/II (Schneider \textit{et al.} 2010), and 297301 objects in the 12th data release for SDSS-III Baryon Oscillation Spectroscopic Survey (BOSS) (Alam \textit{et al.} 2015). The scale of the sample makes the detection of absorptions for the inflows much more available. In fact, Hall \textit{et al.} (2002) reported a few quasars with BAL troughs extending redshifted to the objects' rest frames. The most interesting one is SDSS J112526.12+002901.3 (hereafter J1125+0029), which shows two redshifted absorption troughs of hydrogen Balmer lines at about 70 and $650\ \mathrm{km\ s}^{-1}$, respectively. Both components are also present in the absorptions of metastable \ion{He}{1} $\lambda 3889$, while very strong \ion{Mg}{2} and UV \ion{Fe}{2} absorptions are found at around the same velocity although their exact profiles are unclear due to the overlapping. The rarely detected Balmer and \ion{He}{1}* absorptions, combined with low-ionization metal lines like \ion{Fe}{2}, are powerful diagnostics to the physical conditions of absorbing medium (Leighly \textit{et al.} 2011; Ji \textit{et al.} 2012; Liu \textit{et al.} 2015). Therefore J1125+0029 gives a suitable example to investigate the nature of the redshifted absorption line system.

This paper is organized as follows. In \S\ref{Observation} we describe the data of repeated spectroscopic observations including the SDSS-I/II, BOSS, and MMT. In \S\ref{Measurement} we measure the Balmer and \ion{He}{1}* absorption lines in the spectra using the curve of growth (COG) analysis.
The physical conditions of the absorbing medium are estimated using the photoionization simulations in \S\ref{Photoionization}, which provide the basis for the discussion about the origin of the absorption line system in \S\ref{Discussion}. And finally we give a brief summary in \S\ref{Summary}. The flux calibration and variability of the spectroscopic data is discussed in Appendix. Throughout this paper we assume a cosmology with $H_{0}=70\ \mathrm{km\ s}^{-1}\mathrm{Mpc}^{-1}$, $\Omega_{\mathrm{M}}=0.3$ and $\Omega_{\mathrm{\Lambda}}=0.7$.

\section{The Observations}\label{Observation}

\begin{figure*}
\includegraphics[width=\textwidth]{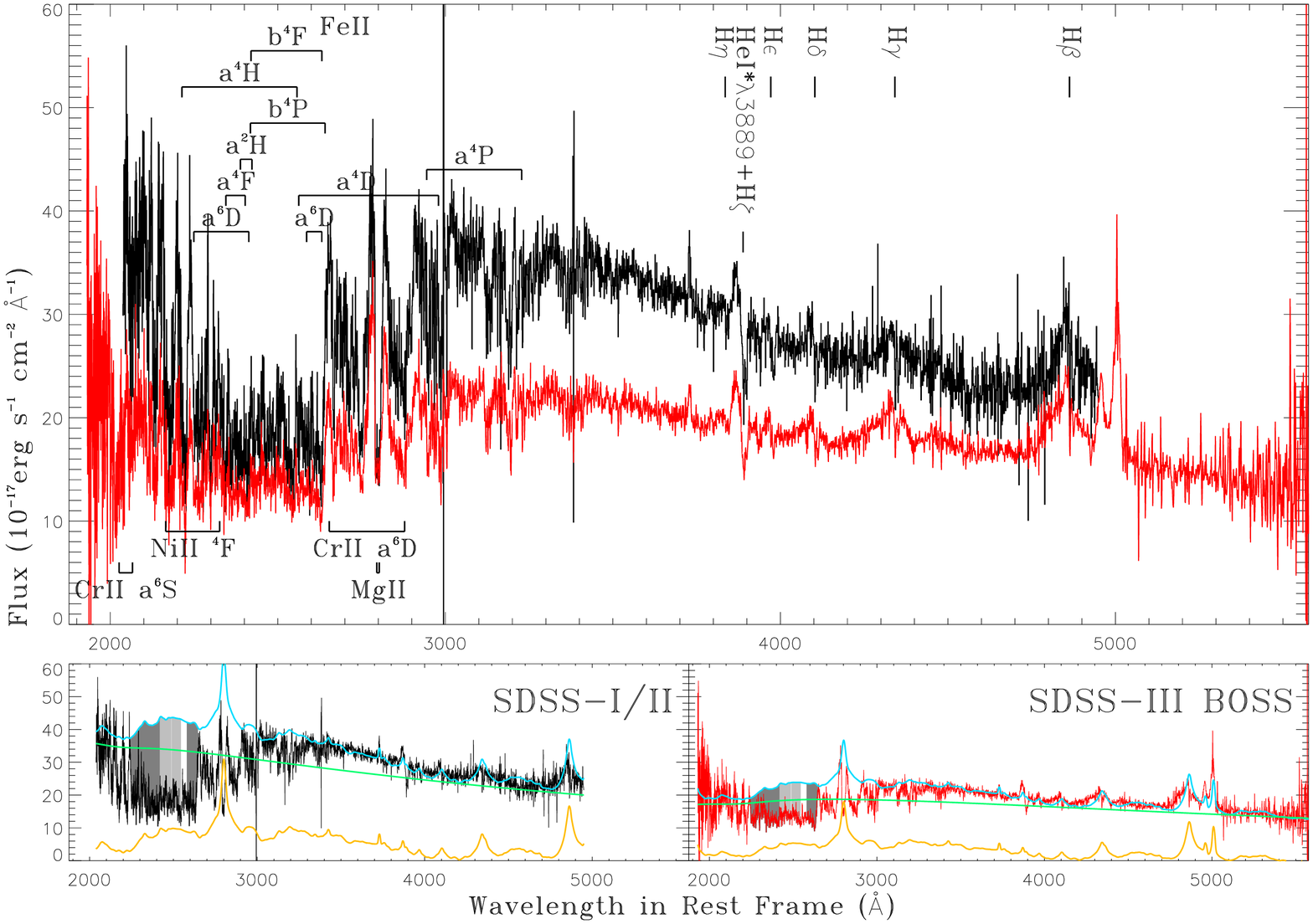}
\caption{Top panel: the SDSS-I/II (black line) and the BOSS (red line) spectra of J1125+0029 corrected for Galactic extinction. The absorptions of Balmer series, \ion{He}{1}*, \ion{Mg}{2}, \ion{Cr}{2}, \ion{Ni}{2}, and the lowest eight terms of Fe$^+$ accounting for the overlapping UV \ion{Fe}{2} troughs are labeled. Bottom panels: the cyan lines are the SDSS quasar composite spectrum (Vanden Berk \textit{et al.} 2001) that can best-fit the observation longward of 3770 \AA in the quasar's rest frame, reddened with SMC-type extinction curve (Gordon \textit{et al.} 2003), for the SDSS-I/II (left) and BOSS (right). The green lines show the reddened power-law continuum, and the orange lines below are the emission of the SDSS composite spectrum. The wavelengths dominated by the ground term of \ion{Fe}{2} are highlighted by the dark gray areas, and those dominated by the terms higher than b$^4$P are highlighted by the light gray areas.\label{SED}}
\end{figure*}

J1125+0029 was observed as a quasar candidate for spectroscopy in SDSS-I/II (York \textit{et al.} 2000) on UT 2000 March 11 and SDSS-III BOSS (Dawson \textit{et al.} 2013) on UT 2011 January 14. The SDSS-I/II spectrum has a resolution of $R \approx 1800$, covering a wavelength range from 3800 to 9200 \AA. To reduce the contamination from the sky line subtraction residual on H$\gamma$ and H$\beta$ peaks, we employ the sky-residual subtracted SDSS DR7 spectrum{\footnote{The spectrum can be accessed through the Johns Hopkins University SDSS server. http://www.sdss.jhu.edu/skypca/spSpec}} post-processed by Wild \& Hewett (2005). The BOSS spectrum covers a wider wavelength range from 3600 to 10000 \AA. It seems the residual of sky line subtraction at the red end of the BOSS spectrum has little effect on our measurement. The spectroscopic data are then corrected for Galactic extinction using the mean extinction curve in Fitzpatrick \& Massa (2007), with selective extinction $E(B-V)=0.031$ in the Galactic dust map of Schlegel \textit{et al.} (1998). The narrow [\ion{O}{2}] $\lambda 3728$ emission presents a good measure of the systemic redshift (Hewett \& Wild 2010, Shen \textit{et al.} 2016). By fitting a Gaussian profile to the [\ion{O}{2}] $\lambda 3728$ line, we suggest a systemic redshift of $0.8632\pm 0.0002$. In Figure \ref{SED}, the SDSS-I/II and BOSS spectra are presented together.

The overall fluxes of the two spectra are significantly different. The fluxes of SDSS-I/II spectrum are about 60\% higher than those of BOSS spectrum. The cause could be the variability and the flux calibration problems in the reduction of the spectra. A detailed discussion about the variability is presented in Appendix.

Despite the great difference in the levels of fluxes, numerous similar absorption features can be identified in both spectra. In the part of spectra shortward of rest-frame 3000 \AA, a vast number of overlapping absorption troughs are the most prominent spectral features. Few windows are left absorption free in this part. The majority can be ascribed to the absorptions from \ion{Fe}{2} multiplets UV1 at $\sim 2600\ \mathrm{\AA}$ up to UV168 at $\sim 2200\ \mathrm{\AA}$. Others are identified as absorptions of \ion{Mg}{2} ($\sim 2800\ \mathrm{\AA}$), \ion{Cr}{2} ($\sim 2060\ \mathrm{\AA}$, $\sim 2665\ \mathrm{\AA}$ and $\sim 2850\ \mathrm{\AA}$), and \ion{Ni}{2} (2160 to $2325\ \mathrm{\AA}$).

\begin{figure}
\includegraphics[width=0.6\columnwidth]{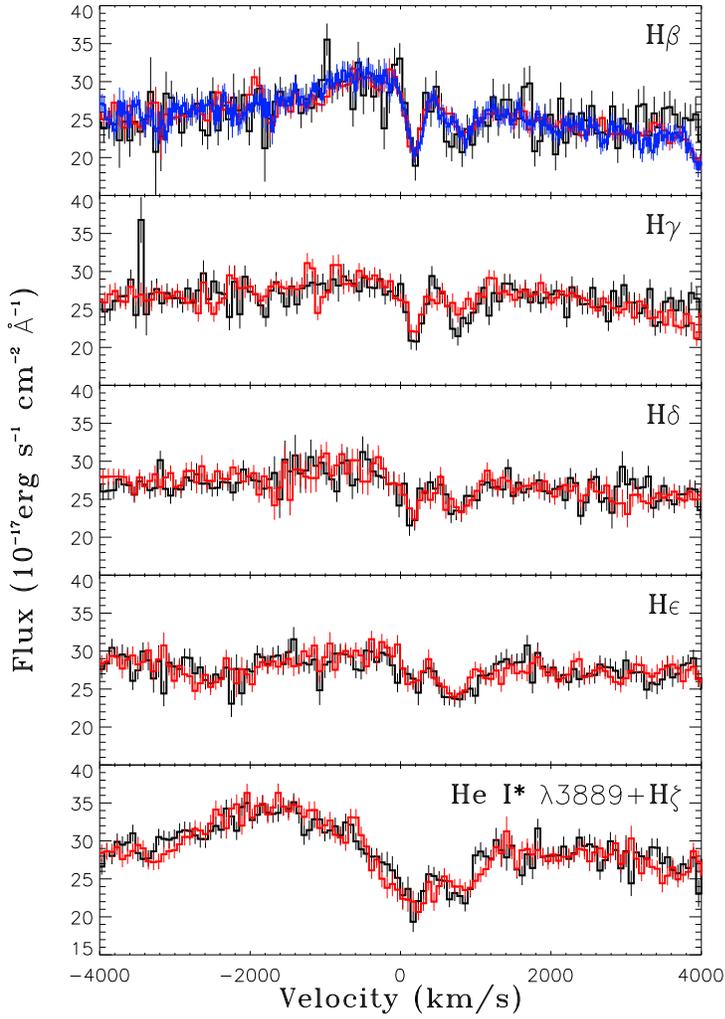}
\caption{The SDSS (black line), BOSS (red line) and MMT (blue line) spectra around H$\beta$, H$\gamma$, H$\delta$, H$\epsilon$, and \ion{He}{1}* $\lambda 3889$ blended with H$\zeta$ are plotted in velocity with respect to the quasar's rest frame. The BOSS and MMT fluxes are scaled to match the SDSS spectrum in the sections $-4000$ to $-200\ \mathrm{km\ s}^{-1}$ and 1400 to $3400\ \mathrm{km\ s}^{-1}$. All these lines have two isolated components.\label{abs_var}}
\end{figure}

The rarely detected Balmer series and \ion{He}{1}* $\lambda 3889$ absorptions are also present in both spectra. In Figure \ref{abs_var} we show the spectra around these absorptions in the velocity with respect to the quasar's rest frame, with the BOSS spectrum being scaled so that the fluxes in the absorption free sections $-4000$ to $-200\ \mathrm{km\ s}^{-1}$ and 1400 to $3400\ \mathrm{km\ s}^{-1}$ neighboring the troughs match those in the SDSS-I/II spectrum. It could thus be clearly seen that in both observations, these absorptions show two redshifted velocity components centered at $\sim 70$ and $650\ \mathrm{km\ s}^{-1}$, respectively, while no detectable changes in these central velocities are found. And the changes in the relative depth of these absorption profiles are much smaller compared with the difference in the overall fluxes.

To further monitor the absorption variability, a near-infrared (NIR) spectrum was obtained on the MMT red channel echellette using an $180\arcsec \times 1\arcsec$ slit on UT 2012 March 1. The exposure time is 1200 s, and the typical seeing is $0\arcsec .8$. The resolution is $R\approx 3270$, slightly higher than the SDSS data. We use the standard IRAF package to extract the 1-D spectrum and the fluxes are carefully calibrated. The spectrum covers H$\beta$ and narrow [\ion{O}{3}] $\lambda\lambda 4960,5008$ emission. The MMT spectrum around the H$\beta$ absorption is also plotted in the top panel of Figure \ref{abs_var}, scaled in the same way as the BOSS spectrum. The profile seems in good agreement with the BOSS observation, indicating that no variation of absorption can be detected from 2011 to 2012, about 7 months in the quasar's rest frame.

\section{Absorption Measurement}\label{Measurement}

In Shi \textit{et al.} (2016), we study the Balmer BAL quasar SDSS J125942.80+121312.6 (hereafter J1259+1213) in detail with the help of photoionization models. Many similarities can be found in the absorption features between J1125+0029 and J1259+1213. First, in both spectra we observe the overlapping troughs of low-ionization metal lines like UV \ion{Fe}{2} and \ion{Mg}{2} as well as the isolated Balmer lines and \ion{He}{1}*. Second, the UV \ion{Fe}{2} troughs between rest-frame 2400 and 2550 \AA, of which the majority are dominated by transitions from the terms with exciting energy $E_{\mathrm{ex}}>2.5\ \mathrm{eV}$, show relative depths of $\sim 0.5$. This implies a highly excited high column density \ion{Fe}{2} absorber, in which the resonant absorptions should be saturated, while large residual fluxes can be measured under the UV1 and UV2,3 multiplets from the ground term. Using the SDSS quasar composite (Vanden Berk \textit{et al.} 2001) intrinsically reddened by the SMC-type extinction curve (Gordon \textit{et al.} 2003) to match the fluxes in the absorption free windows around rest-frame 2100 \AA and longward of 3700 \AA, we find that the residuals under the UV1 and UV2,3 troughs are considerably larger than the UV \ion{Fe}{2} emission bump of the composite. Third, the H$\beta$ troughs are only slightly stronger than the H$\gamma$ troughs, given that the oscillator strength of H$\beta$ is more than twice as large as the oscillator strength of H$\gamma$.

In Shi \textit{et al.} (2016), we find that a high density and high column density gaseous medium can account for the BAL troughs of all observed ions in J1259+1213, including Balmer lines, \ion{He}{1}*, \ion{Mg}{2} and \ion{Fe}{2}. The absorbing medium covers part of the continuum source and little of the broad emission-line region (BELR), as the latter is two orders of magnitudes larger in size. Accordingly we also suppose that the low-ionization metal absorption lines in J1125+0029 are originated from the same medium as the redshifted Balmer lines and \ion{He}{1}* $\lambda 3889$, and this medium only obscures a fraction of the continuum source.

\begin{figure}
\includegraphics[width=\textwidth]{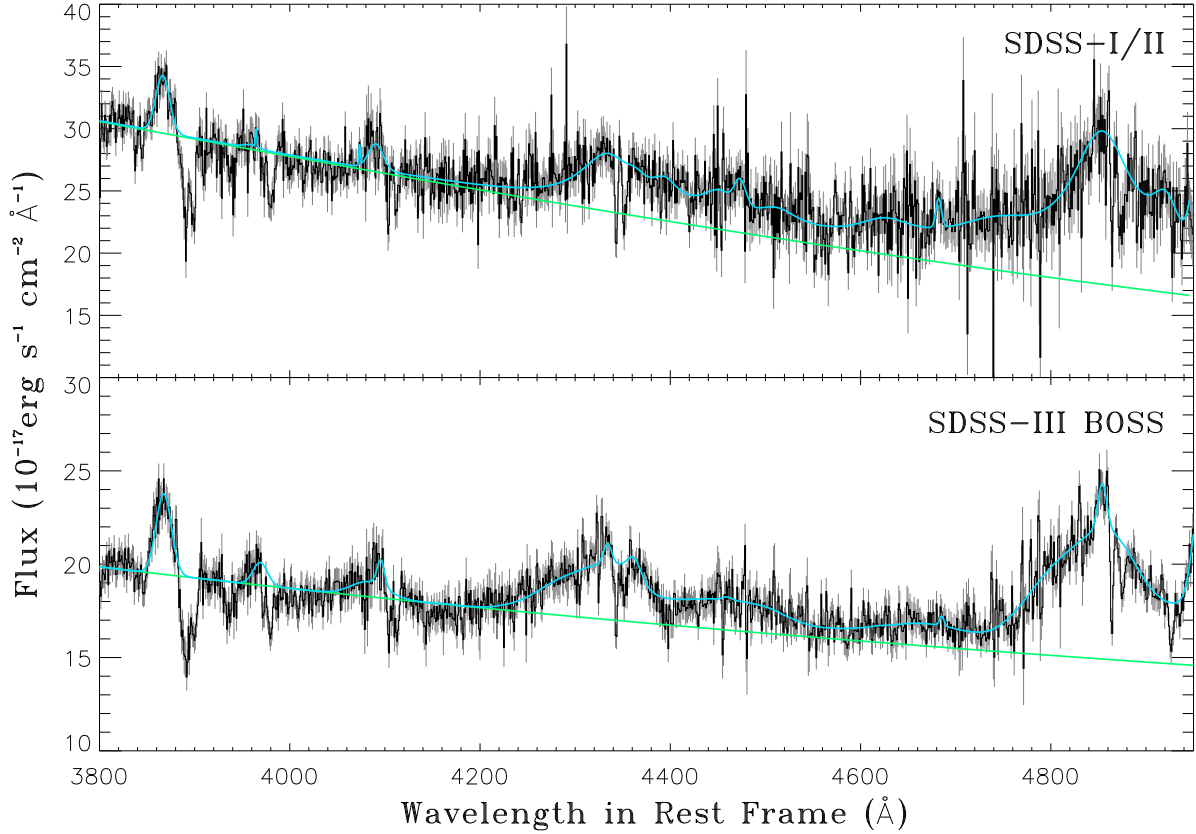}
\caption{The SDSS-I/II and BOSS spectra are modeled following Dong \textit{et al.} (2008). The broad H$\beta$ and H$\gamma$ emissions are fitted using three Gaussian profiles. The underlying power-law continuum, in green line, is determined on the fluxes at around 3550, 3820, 4020, 4160 and $4560\ \mathrm{\AA}$ in the quasar's rest frame, which are regarded as emission-free. The cyan lines present the overall model spectra including power-law continuum, broad and narrow Balmer emission, narrow [\ion{O}{2}], [\ion{O}{3}], [\ion{Ne}{3}] emissions and optical \ion{Fe}{2} emissions.\label{emis_model}}
\end{figure}

According to such assumption, to extract the normalized profile for the measurement of absorptions, the emission lines and unabsorbed continuum should be properly modeled. We would first remove the contribution of emission lines from the spectrum and then divide the residual by the model continuum. We fit the spectrum longward of rest-frame 3500 \AA following the steps described by Dong \textit{et al.} (2008) with small modification. The continuum is modeled using a single power-law continuum multiplied by the SMC-type extinction law (Gordon \textit{et al.} 2003). The continuum windows assumed nearly emission-free are rest-frame 3540 to 3560, 3810 to 3830, 4005 to 4035, 4150 to 4170 and 4550 to $4570\ \mathrm{\AA}$. The narrow emission lines such as [\ion{O}{2}] $\lambda 3728$ and [\ion{Ne}{3}] $\lambda 3868$ are modeled with a single Gaussian profile. The broad Balmer emissions H$\beta$ and H$\gamma$ are assumed to have the same redshift and profile, and modeled using three Gaussian profiles. Other weak broad emissions are modeled using one Gaussian profile. The fitting results are shown in Figure \ref{emis_model}. The measured value of full width at half maximum (FWHM) for the broad H$\beta$ emission in the SDSS-I/III spectrum is $7230\ \mathrm{km\ s}^{-1}$.

\begin{figure}
\includegraphics[width=0.6\columnwidth]{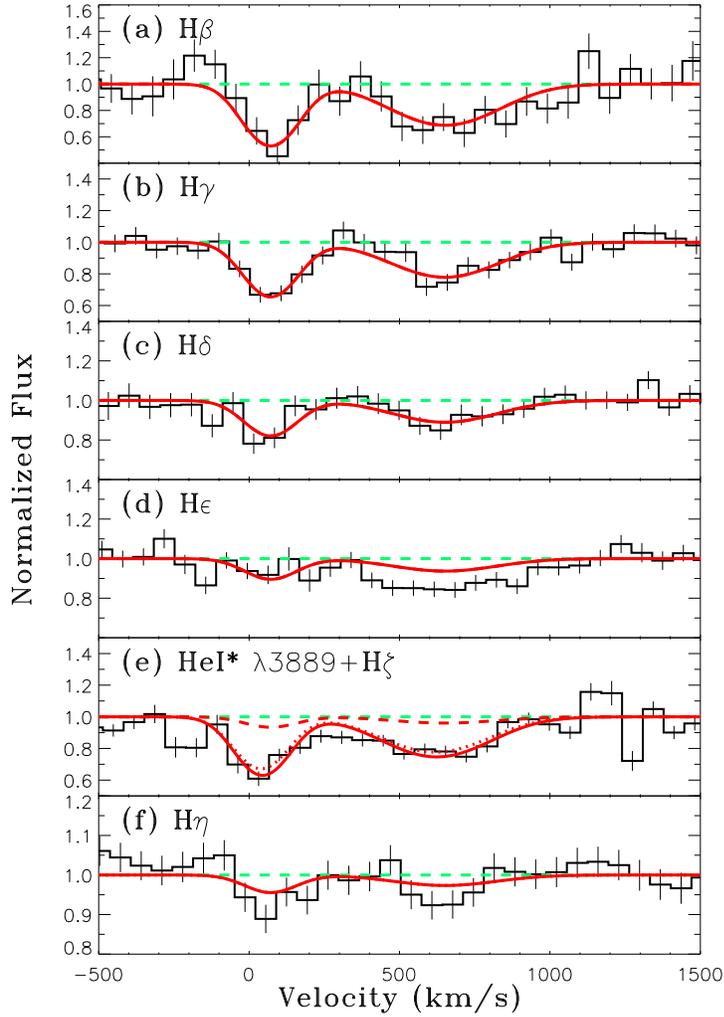}
\caption{Normalized fluxes for H$\beta$, H$\gamma$, H$\delta$, H$\epsilon$, \ion{He}{1}* $\lambda 3889$ plus H$\zeta$, and H$\eta$ in velocity. All these absorptions are fitted using two Gaussian profiles. In panel (e), the dotted red line represents the profile of \ion{He}{1}* $\lambda 3889$, and the dashed line for $H\zeta$. The best fits for the centroidal velocity shifts are $72\pm 39$ for the blue component and $651\pm 41\ \mathrm{km\ s}^{-1}$ for the red component. The FWHMs are $199.4\pm 16.4$ and $398.6\pm 32.6\ \mathrm{km\ s}^{-1}$, respectively.\label{abs_prof}}
\end{figure}

The emission model subtracted spectrum around H$\beta$, H$\gamma$, H$\delta$, H$\epsilon$, \ion{He}{1}* $\lambda 3889$ plus H$\zeta$, and H$\eta$ is normalized using model power-law continuum and the result for SDSS-I/II observation is plotted in Figure \ref{abs_prof}. Since the \ion{He}{1}* and Balmer absorption lines show similar profiles, we assume the \ion{He}{1}* and Balmer absorbers share the same kinematic structure and can fit two Gaussians to the normalized flux of all these lines simultaneously. The velocity shifts with respect to the QSO's rest frame are $72\pm 39$ and $651\pm 41\ \mathrm{km\ s}^{-1}$ for the two components respectively, including the uncertainty of systemic redshift. The FWHMs are $199.4\pm 16.4\ \mathrm{km\ s}^{-1}$ for the blue component and $398.6\pm 32.6\ \mathrm{km\ s}^{-1}$ for the red component. Spanning $\sim 1200\ \mathrm{km\ s}^{-1}$, the whole absorption line system can be classified as mini-BAL. Since the resolution of SDSS-I/II is $\sim 1800$, the $\mathrm{FWHM}_{\mathrm{inst}}$ for the instrumental profile is $\sim 167\ \mathrm{km\ s}^{-1}$. Employing the simple relation $\mathrm{FWHM}_{\mathrm{obs}}^{2}=\mathrm{FWHM}_{\mathrm{true}}^{2}+\mathrm{FWHM}_{\mathrm{inst}}^{2}$, the intrinsic $b$-values for two components are 64 and $215\ \mathrm{km\ s}^{-1}$, respectively.

\begin{figure*}
\includegraphics[width=\columnwidth]{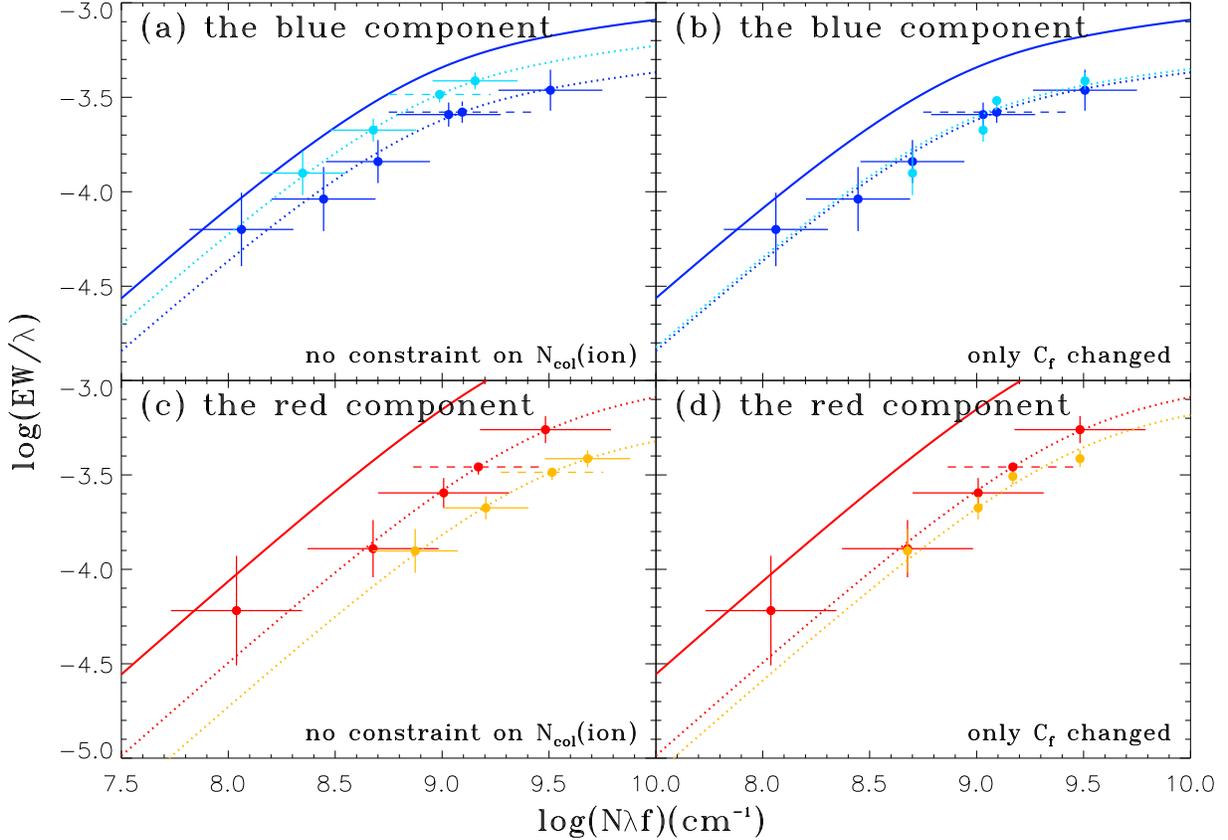}
\caption{The COG analysis for the Balmer and \ion{He}{1}* absorption lines. The solid lines are the COGs specialized by measured $b$-values for the blue and red components, respectively. For the blue component (panels (a) and (b)), the blue data points and the blue dotted lines are the measured lines and the apparent COGs reduced by $C_f$ for the SDSS-I/II observation, and the cyan ones for the BOSS observation, while for the red component (panels (c) and (d)) the red ones are for the SDSS-I/II observation and the orange ones for the BOSS observation. By fitting the apparent COGs to the line measurements, we can derive the optimal values for $N_{\mathrm{col}}(\mathrm{H}^{0}_{n=2})$, $N_{\mathrm{col}}(\mathrm{He}^{0}_{2^{3}\mathrm{S}})$, and $C_f$. The vertical bars associated with the data points are the errors of EWs. The horizontal bars show the fitting errors for the ionic column densities. The data points with solid error bars represent the measurements for Bamler lines, while those with dashed error bars represent the \ion{He}{1}* $\lambda 3889$. In the panels (a) and (c), these values are fitted freely. In the panels (b) and (d), the ionic column densities are supposed unchanged between the two spectroscopic observations.\label{COG}}
\end{figure*}

The equivalent widths (EWs) and the $1\sigma$ uncertainties can be measured directly using the normalized fluxes and the fluctuations in the wavelength range defined by the Gaussian profiles. Since the absorbing medium only covers part of the continuum source, the apparent EWs are the reduced values of the true EWs by a factor of $C_f$, $\mathrm{EW_{app}}=\mathrm{EW_{true}}\times C_f$, where $C_f\leq 1$ is the covering factor. The values of covering factors and ionic column densities $N_{\mathrm{col}}(\mathrm{ion})$ can be derived using the COG analysis. In Figure \ref{COG}, for each component with the known $b$-value, the solid line shows the COG, while the dotted line shows the apparent EWs predicted by COG given the covering factors, $\log \mathrm{EW_{app}}/\lambda=\log \mathrm{EW_{true}}/\lambda+\log C_f$. For those measured absorption lines (represented using filled circles), the ordinated values show the measurements for apparent EWs, while the abscissa values, $N_{\mathrm{col}}(\mathrm{ion}) f \lambda_{\mathrm{rest}}$, would be determined by the ionic column densities $N_{\mathrm{col}}(\mathrm{ion})$. Appointing the unknown $C_f$, $N_{\mathrm{col}}(\mathrm{H}^{0}_{n=2})$, and $N_{\mathrm{col}}(\mathrm{He}^{0}_{2^3 \mathrm{S}})$ as adjustable parameters, we can get the optimal values and $1\sigma$ uncertainties for them by fitting the COG to the measured data points.

Fitting the measured EW values with the COG for the two spectra respectively, for the blue component we have $C_f=0.53\pm 0.18$, $\log N_{\mathrm{col}}(\mathrm{H}^{0}_{n=2})(\mathrm{cm}^{-2})=14.74\pm 0.24$, and $\log N_{\mathrm{col}}(\mathrm{He}^{0}_{2^3 \mathrm{S}})(\mathrm{cm}^{-2})=14.69\pm 0.34$ in the SDSS-I/II observation, and $C_f=0.73\pm 0.20$, $\log N_{\mathrm{col}}(\mathrm{H}^{0}_{n=2})(\mathrm{cm}^{-2})=14.39\pm 0.20$, and $\log N_{\mathrm{col}}(\mathrm{He}^{0}_{2^3 \mathrm{S}})(\mathrm{cm}^{-2})=14.59\pm 0.24$ in the BOSS observation. For the red component, $C_f=0.37\pm 0.23$, $\log N_{\mathrm{col}}(\mathrm{H}^{0}_{n=2})(\mathrm{cm}^{-2})=14.72\pm 0.31$, and $\log N_{\mathrm{col}}(\mathrm{He}^{0}_{2^3 \mathrm{S}})(\mathrm{cm}^{-2})=14.77\pm 0.30$ in the SDSS-I/II observation, and $C_f=0.22\pm 0.06$, $\log N_{\mathrm{col}}(\mathrm{H}^{0}_{n=2})(\mathrm{cm}^{-2})=14.92\pm 0.20$, and $\log N_{\mathrm{col}}(\mathrm{He}^{0}_{2^3 \mathrm{S}})(\mathrm{cm}^{-2})=15.11\pm 0.24$ in the BOSS observation (see Figure \ref{COG} panels (a) and (c)).

It can be found that for all parameters, the changes from the SDSS-I/II to the BOSS observations are not larger than about $1\sigma$ error. That means the absorption variability between the SDSS-I/II and BOSS observations is not significant. If we suppose the changes of the covering factors as the only cause of absorption variability (with the ionic column densities unchanged), a picture frequently prompted to explain the absorption variability (Hall \textit{et al.} 2002; Zhang \textit{et al.} 2015b), the covering factors are 0.55 and 0.30 for the blue and red components in the BOSS observation, respectively, with the reduced $\chi^2$ still being around 1 (see Figure \ref{COG} panels (b) and (d)). The uncertainties for $C_f$ are reduced to 0.14 and 0.07, respectively. It seems that the assumption that the physical conditions of the absorbing medium remain unchanged is acceptable.

\section{Photoionization Models for the Absorbing Medium}\label{Photoionization}

We use the photoionization code CLOUDY (version 10.00, last described by Ferland \textit{et al.} 1998) to simulate the ionization process, assuming a simple model of slab-shaped geometry, unique density and homogeneous chemical composition of solar values for the absorbing medium. The incident SED applied is the combination of a UV bump described as $\nu^{\alpha_{\mathrm{UV}}}\mathrm{exp}(-h\nu /kT_{\mathrm{BB}})\mathrm{exp}(-kT_{\mathrm{IR}}/h\nu)$ and power-law $a\nu^{\alpha_{\mathrm{X}}}$, incorporated in CLOUDY. This is considered typical for observed AGN continuum. The UV bump is parameterized by UV power-law index $\alpha_{\mathrm{UV}}=-0.5$, and exponentially cut off with temperature $T_{\mathrm{BB}}=1.5\times 10^5\ \mathrm{K}$ at high energy and $kT_{\mathrm{IR}}=0.01\ \mathrm{Ryd}$ at infrared. The power-law component has an index $\alpha_{\mathrm{X}}=-2$ beyond 100 keV, and $-1$ between 1.36 eV and 100 keV. The overall flux ratio of X-ray to optical is $\alpha_{\mathrm{OX}}=-1.4$. The physical conditions of the absorbing medium are characterized by the ionization parameter $U$ at the irradiated surface, the total hydrogen density $n(\mathrm{H})$, and the total hydrogen column density $N_{\mathrm{col}}(\mathrm{H})$ which indicates the thickness of the medium.

The \ion{He}{1}* $\lambda 3889$ absorption is originated from the metastable He$^0$ 2$^3$S level, which in the photoionization dominated medium is populated through the recombination of He$^+$ in the ionized zone. Ji \textit{et al.} (2015) presented a detailed investigation on the \ion{He}{1}* ionization structure using the photoionization simulations. They found that if the medium is thick enough that the ionizing front is well developed, the value of $U$ can solely determine the column density of \ion{He}{1}*. Given $\log N_{\mathrm{col}}(\mathrm{He}^{0}_{2^3 \mathrm{S}})(\mathrm{cm}^{-2})=14.69\pm 0.34$ and $14.77\pm 0.30$ for the blue and red components, we obtain $\log U=-1.9\pm 0.3$ and $-1.8\pm 0.3$ according to the Figure 10 in Ji \textit{et al.} (2015), respectively.

\begin{figure*}
\includegraphics[width=\textwidth]{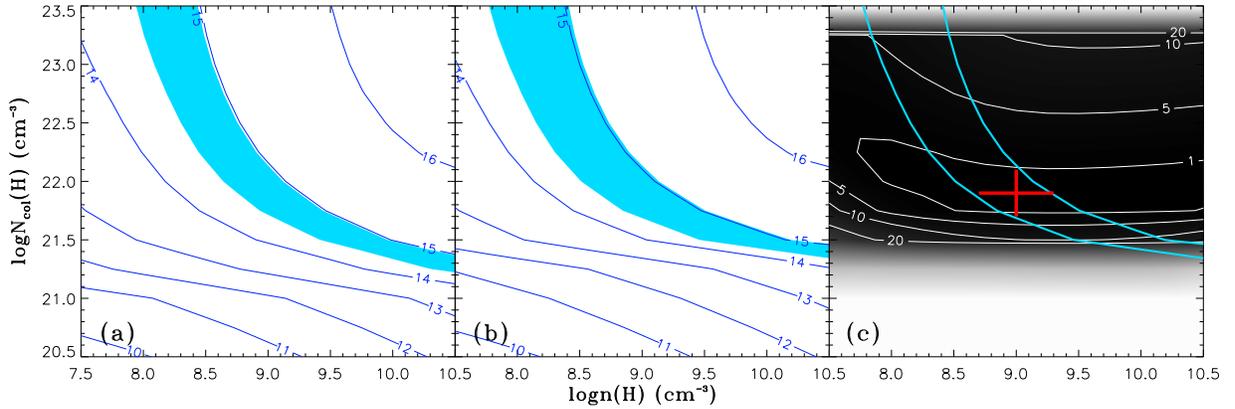}
\caption{The predicted ionic column densities for Balmer lines from the photoionization simulations by CLOUDY with $\log U=-1.9$ for the blue component (panel (a)) and $\log U=-1.8$ for the red component (panel (b)) as functions of $n(\mathrm{H})$ and $N_{\mathrm{col}}(\mathrm{H})$. The numbers labeling the contours are the logarithms of ionic column densities. The cyan areas represent the measured values with $1\sigma$ error for Balmer lines.
In panel (c) we plot the distribution of $\chi^2_{\nu}$ which evaluate the difference between the SDSS-I/II observation and the synthetic model spectra in the overlapping UV \ion{Fe}{2} troughs between rest-frame 2320 and 2780 \AA. The numbers labeling the contours are $\chi^2_{\nu}-\chi^2_{\nu,\mathrm{min}}$. Supposing the physical parameters for the two components are the same, we find the optimal values are $\log n(\mathrm{H})(\mathrm{cm}^{-3})=9\pm 0.3$ and $\log N_{\mathrm{col}}(\mathrm{H})(\mathrm{cm}^{-2})=21.9\pm 0.2$.\label{model_distri}}
\end{figure*}

The hydrogen $n=2$ shell could be populated through a couple of mechanisms, including recombination, collisional excitation, and Ly$\alpha$ resonant scattering. Therefore, Balmer absorption lines can be originated in both ionized and neutral zones, showing more complicated dependence on the density and total column density of the medium as well as the strength of ionizing flux, while the density of $n(\mathrm{H})>10^6\ \mathrm{cm}^{-3}$ is generally required. In Figure \ref{model_distri} panels (a) and (b), we plot the column density of H$^0_{n=2}$ as functions of $n(\mathrm{H})$ and $N_{\mathrm{col}}(\mathrm{H})$ predicted by the photoionization simulations given the values of $U$, and the measured values are highlighted by the colored areas.

To further constrain the physical conditions of the absorbing medium, the measurements for other ions in the same medium are required, such as Fe$^+$. But unlike the case of J1259+1213 where the optical absorption troughs of Fe$^+$ $\lambda 4233$, $\lambda 4924$, $\lambda 5018$, and $\lambda 5169$ from excited Fe$^+$ are presented isolated which can be directly demonstrated associated with Balmer lines and reliably measured, in the spectra of J1125+0029 these lines can hardly be detected due to the poorer S/N. The UV \ion{Fe}{2} absorption troughs are so heavily saturated and blended that we cannot find isolated absorption lines to estimate the column density on any individual level of Fe$^+$. But being aware of the obvious similarity in appearance between these two objects, we guess in J1125+0029 the UV \ion{Fe}{2} troughs also come from the same absorbing gas as Balmer lines. That means each individual \ion{Fe}{2} trough would also has two components with the same central velocities, widths, and covering factors as Balmer lines since Fe$^+$ and H$^0_{n=2}$ are originated from the similar region in the photoionized medium.

Following the method described in Shi \textit{et al.} (2016), we can construct the synthetic model UV \ion{Fe}{2} absorption spectra based on the absorption profile extracted from Balmer lines and the column densities of Fe$^+$ on various levels predicted by the simulations. Since considerable contribution from Fe$^+$ excited levels with $E_{\mathrm{ex}}>2.5\ \mathrm{eV}$ can be identified in the overlapping UV \ion{Fe}{2} troughs, the full 371 levels Fe$^+$ model incorporated in CLOUDY is used in our simulations, including all levels up to 11.6 eV. Comparing the model spectra with the observation, we can use the UV \ion{Fe}{2} absorption features to further constrain the physical conditions of the absorbing medium.

\begin{figure*}
\includegraphics[width=\textwidth]{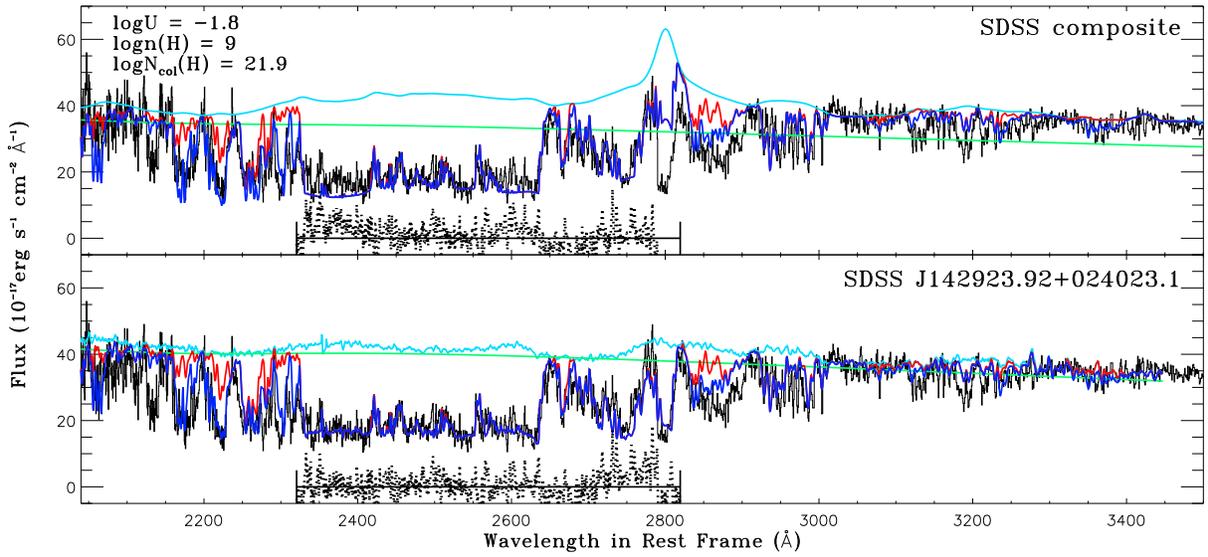}
\caption{The synthetic model spectrum for the optimal photoionization model. The red line represents the profile of \ion{Fe}{2} and \ion{Mg}{2} absorptions, and the blue line includes absorption from \ion{Cr}{2} and \ion{Ni}{2} in addition. The cyan line is the template employed as unabsorbed background radiation to construct the synthetic spectrum, and the green line is the corresponding power-law continuum. In the top panel, reddened SDSS quasar composite is used as unabsorbed template. In the bottom panel, reddened spectrum of SDSS J142923.92+024023.1 is used as template. The residual of the synthetic model spectra is also plotted (the underlying data points).\label{synth_spec}}
\end{figure*}

In Figure \ref{synth_spec} top panel we plot an example of the synthetic model UV \ion{Fe}{2} absorption spectra. The best-fitting reddened SDSS quasar composite shown in Figure \ref{SED} is employed as the unabsorbed template. The reduced $\chi^2_{\nu}$ calculated in the range from rest-frame 2320 to 2780 \AA, where \ion{Fe}{2} contributes to nearly all the absorption features, is used to assess the agreement between the synthetic model spectra and the observation. With no further constraint, we find that the physical parameters leading to the minimum $\chi^2_{\nu}$ are $\log n(\mathrm{H})(\mathrm{cm}^{-3})\sim 9$ and $\log N_{\mathrm{col}}(\mathrm{H})(\mathrm{cm}^{-2})\sim 22$ for both components. Thus, it would be convenient and reasonable to suppose the physical states of the two components are identical. In Figure \ref{model_distri} panel (c), we plot the distribution of $\chi^{2}_{\nu}$ for the models of which the parameters for the blue and red components are the same.

The optimal values for the parameters $n(\mathrm{H})$ and $N_{\mathrm{col}}(\mathrm{H})$ are the values which present the measured value of $N_{\mathrm{col}}(\mathrm{H}^{0}_{n=2})$ and the minimum of $\chi^{2}_{\nu}$, while the $1\sigma$ uncertainty for these parameters are given by the area defined by the $1\sigma$ uncertainty of $N_{\mathrm{col}}(\mathrm{H}^{0}_{n=2})$ and the contour of $\chi^{2}_{\nu,\mathrm{min}}+1$. Thus we have $\log n(\mathrm{H})(\mathrm{cm}^{-3})=9\pm 0.3$ and $\log N_{\mathrm{col}}(\mathrm{H})(\mathrm{cm}^{-2})=21.9\pm 0.2$.

Other low-ionization ions, such as Ti$^+$, Cr$^+$, and Ni$^+$, can also be included in the synthetic model spectra following the same method as Fe$^+$. In Figure \ref{synth_spec} we also plot this complete model for the \ion{Fe}{2} selected optimal models. The inclusion of the absorptions from these ions makes the model better match the observation, with the only exception of \ion{Mg}{2}. The model \ion{Mg}{2} absorption is much shallower than observation. Since we suppose that the absorbing medium covers part of the accretion disk, the fluxes under \ion{Mg}{2} troughs should consist of the fluxes of \ion{Mg}{2} emission and unobscured part of continuum. But the observed fluxes under \ion{Mg}{2} troughs are smaller than \ion{Mg}{2} emission peak in the quasar composite. Such deviation could be ascribed to the difference between the \ion{Mg}{2} emission in the composite and the true \ion{Mg}{2} emission in our source, because there is considerable object to object variation in quasar emission lines. If we suppose the \ion{Mg}{2} emission in J1125+0029 is much weaker than that in the composite, the disagreement can be reduced. Leighly \textit{et al.} (2011) used a dozen of real non-absorption quasars to match the NIR spectrum of FBQS J1151+3822, and chose the best matched one as template to measure the \ion{He}{1}* $\lambda 10830$. Zhang \textit{et al.} (2014) and Liu \textit{et al.} (2015) also developed a similar pair-matching method to improve the estimate of unabsorbed level. We search all SDSS DR7 non-BAL quasar spectra with mean S/N per pixel greater than 15 for the suitable template. The spectrum of SDSS J142923.92+024023.1 showing very weak \ion{Mg}{2} emission best meets our request if it is reddened with $E(B-V)=0.045$. The synthetic spectrum is plotted in the bottom panel of Figure \ref{synth_spec}. Compared with the synthetic spectrum based on the reddened composite, the result is improved not only for the \ion{Mg}{2} doublets but also for the UV \ion{Fe}{2} from rest-frame 2320 to 2640 \AA.

\section{Discussion}\label{Discussion}

The mass of the central SMBH can be estimated according to the relation in Wang \textit{et al.} (2009), assuming the broad emission line region (BELR) is virialized, $\log (M_{\mathrm{BH}}/10^{6}M_{\sun})=(1.39\pm 0.14)+0.5\log (\lambda L_{5100}/10^{44}\ \mathrm{erg\ s}^{-1})+(1.09\pm 0.23)\log (\mathrm{FWHM}(\mathrm{H}\beta)/1000\ \mathrm{km\ s}^{-1})$, where $L_{5100}$ is the $\lambda L_{\lambda}$ at rest-frame 5100 \AA. Thus, we have $M_{\mathrm{BH}}=1.5\times 10^{9}\ M_{\sun}$ with an uncertainty of a factor of $\sim 2.5$ (from the intrinsic scatter $\sim 0.4$ dex for this single-epoch method compared with the results of reverberation mapping, Ho \& Kim 2015). And adopting the correction factor by Runnoe \textit{et al.} (2012), the bolometric luminosity is $L_{\mathrm{bol}}=(8.1\pm 0.4)\times L_{5100}=3.9\pm 0.2\times 10^{46}\ \mathrm{erg\ s}^{-1}$. Assuming an accretion efficiency of 0.1, the mass accretion rate is then $\dot{M}_{\mathrm{BH}}\approx 6.8\ M_{\sun}\ \mathrm{yr}^{-1}$. Extrapolating the best-fitting SDSS composite presented in Figure \ref{SED}, we can derive the continuum flux at 1215.67 \AA. Then given the model incident SED used in simulations and the systemic redshift, the pre-extinction monochromatic luminosity of ionizing continuum at Lyman limit can be roughly estimated. For the SDSS-I/II spectrum, this value is $L_{\nu}(912)=5.0\times 10^{30}\ \mathrm{erg\ s}^{-1}\mathrm{Hz}^{-1}$.

Given the luminosity and SED of the ionizing continuum, we can derive the distance of the absorbing medium to the central engine according to the physical conditions constrained by the photoionization models, as $\frac{L(<912)}{4\pi r_{\mathrm{abs}}^2}=Un(\mathrm{H})c\overline{E_{\mathrm{ph}}(<912)}$. $L(<912)$ is the ionizing luminosity of the continuum source, determined by $L_{\nu}(912)$ and the incident SED used in simulation models. And $\overline{E_{\mathrm{ph}}(<912)}$ is the average energy for all ionizing photons, which also can be evaluated according to the model SED. With $\log U=-1.8\pm 0.3$ and $\log n(\mathrm{H})(\mathrm{cm}^{-3})=9\pm 0.3$ from the optimal photoionization models, the distance of the inner surface of the absorbing medium is $r_{\mathrm{abs}}=4^{+6.6}_{-2.5} \mathrm{pc}$. The listed uncertainty only includes the uncertainties of $U$ and $n(\mathrm{H})$. The uncertain of the AGN ionizing luminosity introduced due to the extrapolation of power-law continuum is more difficult to assess. A change of 100\% for the luminosity can lead to a change of 41\% for the distance, making it a relative minor factor.

In their first paper reported the redshifted absorption line systems in quasars' spectra including J1125+0029, Hall \textit{et al.} (2002) suggested a rotation-dominated disk wind at the phase when the outflow just rises from the accretion disk to explain the redshifted troughs. With the release of BOSS spectra, Hall \textit{et al.} (2013) returned to the issue. Comparing the BOSS and SDSS-I/II spectra of J1125+0029, especially the Mg II troughs, they suggested that the blueshifted part of the Mg II absorption weakened more than the redshifted part, to account for the variability (for more details, see Appendix). The different behavior in the blueshifted and redshifted absorption is believed consistent with the picture of a rotational wind when the cloud moves from the approaching side to the receding side. Such wind is suggested located $1255\ R_{\mathrm{Sch}}$ ($5.5\times 10^{17}\ \mathrm{km\ s}^{-1}$ with an uncertainty of a factor of 2.5) from the SMBH (Murray \& Chiang 1998; Elvis 2000). Another explanation for the redshifted troughs mentioned in Hall \textit{et al.} (2013), the gravitational redshift, requires the absorbing medium to be located at even smaller radii, $\sim 100\ R_{\mathrm{Sch}}$.

The absorbing medium described by our photoionization models seems much more distant than that implied in these explanations. Furthermore, according to the COG analysis in \S\ref{Measurement}, we can estimate the transverse velocities of the absorbing medium by the variations of covering factors for Balmer lines between the SDSS-I/II and BOSS observations. In the simplest picture, these changes ($\Delta C_f=0.02$ for the blue component and 0.07 for the red component) stand for a continuous movement of the medium across our LOS. The radius of the accretion disk where the radiation peaks at 4863 \AA can be evaluated from $\sigma T_{\mathrm{eff}}^{4}=\frac{3GM\dot{M}}{8\pi R^{3}}f(R,a)$ (Eq. 2 in Collin \textit{et al.} 2002). With $k_{\mathrm{B}}T_{\mathrm{eff}}=h\nu$ and the boundary condition $f(R,a)\approx 1$, we obtain $R(4863)\approx 6.1\pm 2.1\times 10^{15}\ \mathrm{cm}$. Considering the uncertainty of $C_f$, although for the blue component the probability that the transverse velocity is smaller than the radial velocity is only 0.56, for the red component the upper limit of transverse velocity at a confidence level of 0.98 is $145\ \mathrm{km\ s}^{-1}$ which is much smaller than the corresponding redshifted velocity. The infall seems a more reasonable explanation.

The distance of the infalling medium is also larger than the radius of H$\beta$ BELR by a factor of $\sim 10$, but how about the torus? The torus has long been suggested as the reservoir of interstellar medium (ISM) feeding the central engine and the direct source of the accretion inflow (Krolik \& Begelman 1988). In recent works, a clumpy model for the torus is required to reproduce the IR SED especially the lack of $10\ \micron$ Si emission feature (Nenkova \textit{et al.} 2002). Such model is supported by Vollmer \textit{et al.} (2004) and Beckert \textit{et al.} (2004) who linked the clumpy torus in AGN with the circumnuclear disk (CND) surrounding the central black hole of our Galaxy which consist of several hundred clouds of gas and dust. The transfer from the thin CND to the thick obscuring torus depends on the accretion rate. The radial accretion flow is now naturally regarded as the result of cloud-cloud collision through which these clouds lost energy and then fall inward. If the falling clouds are located a bit faraway from the equatorial plane, they could be observed in the foreground of the accretion disk since at such view angles the disk would not be severely obscured due to the the relative low local filling factor.

The inner radius of the torus used to be approximated as the evaporation radius, $R_{\mathrm{evap}}=1.3L_{\mathrm{UV,46}}^{1/2}T_{1500}^{-2.8}\ \mathrm{pc}\approx 1.3\ \mathrm{pc}$ following Barvainis 1987 for our object, where $L_{\mathrm{UV,46}}$ is the UV luminosity in unit of $10^{46}\ \mathrm{erg\ s}^{-1}$ estimated using $\lambda L_{\lambda}(1450)$, and $T_{1500}$ is the grain evaporation temperature in unit of 1500 K which is $\sim 1$. By comparing $R_{\mathrm{evap}}$ with the results of reverberation mapping, Kishimoto \textit{et al.} (2007) suggested the inner radius was overestimated in this way by a factor of $\sim 3$. Kawaguchi \& Mori (2011) argued the inner radius should increase with the view angle of the torus due to the anisotropic illumination of the disk, to explain the intrinsic scatter in reverberation mapping (Reminding in J1125+0029, the SED is intermediately reddened by the dust associated with the object, implying that the LOS is not very close to the equatorial plane of the torus). Infrared interferometry, another direct radius measurement, presents the results a factor of 2 larger than those from reverberation mapping (Koshida \textit{et al.} 2014). Thus to date the uncertainty of the inner radius estimate is also quite large, comparable to the uncertainty of our photoionization models. However we can still conclude that the absorbing medium is not far away from the inner surface of the torus.

The infalling absorbing medium is estimated close to the inner surface of the torus, and the density of the medium is found similar to the gas density of the torus. Therefore we strongly suppose the redshifted absorption line system in J1125+0029 representing the accretion flow originated from the torus as our LOS toward the continuum source lying through the low-density part of the clumpy torus. The relation between the two components of the absorbing medium remains unclear. Since the physical conditions and the distance of both components are almost the same, they might reflect the approaching and receding parts of a spinning cloud in the accretion flow with the centroidal infalling velocity $v_{\mathrm{infall}}\approx 350\ \mathrm{km\ s}^{-1}$.

The mass of the absorbing medium is $M_{\mathrm{abs}}=\mu m_{\mathrm{p}}N_{\mathrm{col}}(\mathrm{H})S$, where $\mu$ is the mean atomic mass per proton, $m_{\mathrm{p}}$ is the mass of proton, and $S$ is the projection area. Assuming the observed absorbing medium is typical of all accretion clouds and these clouds are uniformly distributed at the inner surface of torus, we can estimate the mass inflow rate as $\dot{M}_{\mathrm{inflow}}\approx \pi R_{\mathrm{evap}}^2 C_{f,\mathrm{torus}} \mu m_{\mathrm{p}}N_{\mathrm{col}}(\mathrm{H})/t_{\mathrm{infall}}$, where $C_{f,\mathrm{torus}}\approx 0.6$ is the global covering factor of the torus (Lawrence \& Elvis 2010), $\pi R_{\mathrm{evap}}^2 C_{f,\mathrm{torus}}\approx \sum_i S_i$, and $t_{\mathrm{infall}}$ is the infalling timescale of the accretion clouds. If we approximate that $t_{\mathrm{infall}}=R_{\mathrm{evap}}/v_{\mathrm{infall}}$, $\dot{M}_{\mathrm{inflow}}\approx 0.43\ M_{\odot}\ \mathrm{yr}^{-1}$, too small compared with $\dot{M}_{BH}$. Reminding that $v_{\mathrm{infall}}$ is the radial velocity at the initial stage of infall, we think $t_{\mathrm{infall}}$ is overestimated in this way. Since the accretion inflow lies close to the equatorial plane, the radiation from the central engine would be obscured which reduces the radiation pressure, we can use the free fall timescale $t_{\mathrm{ff}}=(R_{\mathrm{evap}}^3/GM_{\mathrm{BH}})^{1/2}$ as $t_{\mathrm{infall}}$. Thus, $\dot{M}_{\mathrm{inflow}}\approx 2.8\ M_{\odot}$ is of the same magnitude as $\dot{M}_{BH}$. And in reality, the properties of clouds in the inflow would be dependent on the height. In J1125+0029, our LOS maybe only passes trough the outskirt of the torus. The closer to the equatorial plane the clouds are, the denser they would be. Thus, the mass inflow rate could be even higher.

\section{Summary}\label{Summary}

The redshifted absorption line systems are rarely detected features in the quasar spectra. A couple of theoretical pictures were prompted to explain the phenomena, but no decisive conclusion has been achieved, unless the physical conditions and the spacial structure of the absorbing medium are available. In the SDSS-I/II and BOSS spectra of quasar J1125+0029, a redshifted absorption line system is identified including lines of hydrogen Balmer series, metastable \ion{He}{1}, \ion{Mg}{2}, \ion{Fe}{2}, \textit{et al.}. These lines are powerful diagnostics to the physical properties of the absorbing medium. Performing a careful measurement of the ionic column densities and covering factors for Balmer and \ion{He}{1}* lines and using the photoionization simulations, we find that the medium is located $\sim 4\ \mathrm{pc}$ away from the central engine and its motion is dominated by infall. Since the distance is consistent with the radius of the inner surface of torus and the physical conditions of the medium are also similar to the torus, we suggest this absorption line system as a candidate for the accretion inflow originated from the torus which fuel the SMBH.

\acknowledgments

This work is supported by the National Basic Research Program of China (the 973 Program 2013CB834905) and the National Natural Science Foundation of China (NSFC-11421303 and 11473025). This research uses data obtained through the Telescope Access Program (TAP), which has been funded by the Strategic Priority Research Program The Emergence of Cosmological Structures (Grant No. XDB09000000), the National Astronomical Observatories, the Chinese Academy of Sciences, and the Special Fund for Astronomy from the Ministry of Finance.

\appendix

\section{The Variability}\label{Variability}

The SDSS-I/II and the BOSS spectra show great difference in the fluxes. In general, the fluxes of SDSS-I/II are about 60\% higher than those of BOSS. Since the strong rapid variability is common for quasars, the difference could reflects the intrinsic variability of luminosity of the object from the SDSS-I/II observation to the BOSS observation (5.8 years in the quasar's rest frame). On the other hand, the spectrophotometric calibration errors in BOSS are reported larger than in SDSS-I/II (Margala \textit{et al.} 2015), and this systematic is wavelength dependent. On average, the miscalibration of the BOSS spectra accounts for a $\sim 19\%$ excess at 3600 \AA and a $\sim 24\%$ decrement at 10000 \AA with a smooth transition between.

\begin{figure*}
\includegraphics[width=\textwidth]{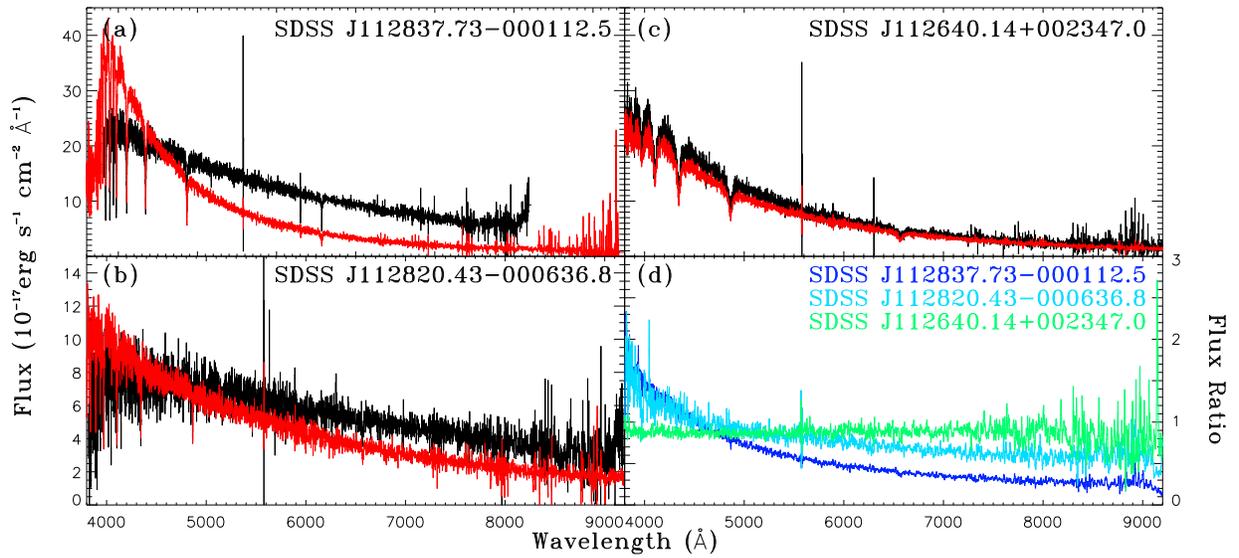}
\caption{Panels (a)-(c): the SDSS-I/II and BOSS spectra of three stars spectroscopically observed on the same plates as J1125+0029. Panel (d): the flux ratios of the SDSS-I/II spectra to the BOSS spectra for these stars. The behavior of the ratios differ from object to object.\label{flux_ratio}}
\end{figure*}

Dawson \textit{et al.} (2013) described the reduction process for the BOSS spectroscopic data. The process is performed independently for targets observed on different fibers, thus the systematics on the objects of the same plate may be different. We find 6 objects observed repeatedly on the same SDSS-I/II and BOSS plates as J1125+0029. Three of them are main sequence stars of which the SED are highly invariable. Assuming the flux calibration for the SDSS-I/II spectra is reliable, we can use these objects to assess the fiber-to-fiber difference of the flux calibration errors for the BOSS data. In Figure \ref{flux_ratio}, we plot the SDSS-I/II and BOSS spectra for these stars, and also the ratio of the BOSS fluxes to the SDSS-I/II fluxes. For two objects, the ratio varies as functions of the wavelength, with the flux shortward of $\sim 4500\ \mathrm{\AA}$ being overestimated and the flux longward of $\sim 4500\ \mathrm{\AA}$ being underestimated which is consistent with the conclusion of Margala \textit{et al.} (2015). But the object-to-object variation is also remarkable. For SDSS J112837.73-000112.5 the ratio declines from $\sim 1.7$ at 3800 \AA to $< 0.5$ at 9200 \AA, while for SDSS J112640.14+002347.0 the ratio is $\sim 0.9$ regardless of the wavelength. Thus the systematics for specified object can hardly be corrected according to the standard stars on the same plate.

\begin{figure*}
\includegraphics[width=\textwidth]{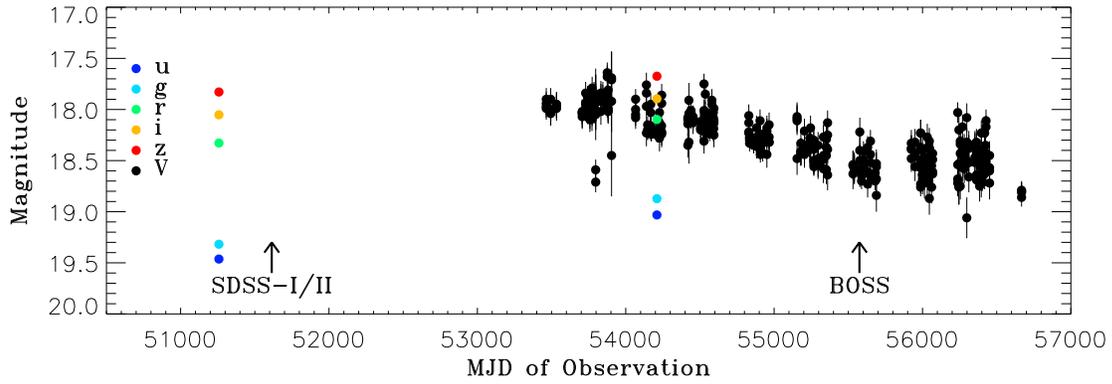}
\caption{The photometric data for J1125+0029. The colored data points show the results of multi-bands SDSS photometry, while the black data points show the light curve of $V$-band Catalina survey. The dates the SDSS-I/II and BOSS spectra being obtained are labeled. It seems the BOSS spectrum cannot be calibrated using any SDSS photometric data.\label{light_curve}}
\end{figure*}

J1125+0029 was monitored photometrically by the $V$-band Catalina survey{\footnote{http://nesssi.cacr.caltech.edu/DataRelease/}} from 2005 April to 2014 January (see Figure \ref{light_curve}). The SDSS images for the object were taken at 1999 March 21 and 2007 April 20. And the BOSS spectrum was obtained 2.0 years after the latest SDSS image observation in the quasar's rest frame. Since the object seems faded steadily in this time interval according to the Catalina light curve, the BOSS spectrum cannot be calibrated using any set of SDSS photometry.

\begin{figure*}
\includegraphics[width=\textwidth]{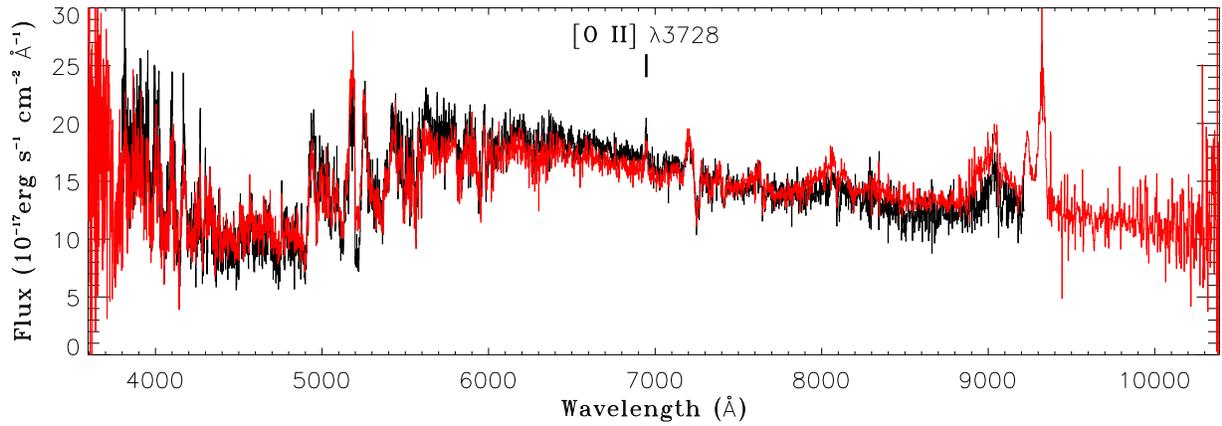}
\caption{Using the narrow emission lines which are nearly invariable to check the flux calibration errors for BOSS data. In this plot the BOSS spectrum (red line) is multiplied by a scaling factor of 1.61 derived from the [\ion{O}{2}] $\lambda 3728$ emission to be compared with SDSS observation (black line).\label{nl_calib}}
\end{figure*}

The narrow [\ion{O}{2}] $\lambda 3728$ emission detected in both spectra is believed not vary significantly in short term, thus presents a useful tool to check the flux calibration problem. The measured strength of [\ion{O}{2}] $\lambda 3728$ in the SDSS-I/II spectrum is 61\% larger than that in the BOSS spectrum. In Figure \ref{nl_calib} we plot the BOSS spectrum multiplied by a scaling factor of 1.61 to be compared with the SDSS-I/II spectrum. Although the details of wavelength-dependent calibration error is still unknown, at least we now realize that the variability of luminosity is not as great as it looks like at the first sight.

\begin{figure*}
\includegraphics[width=\textwidth]{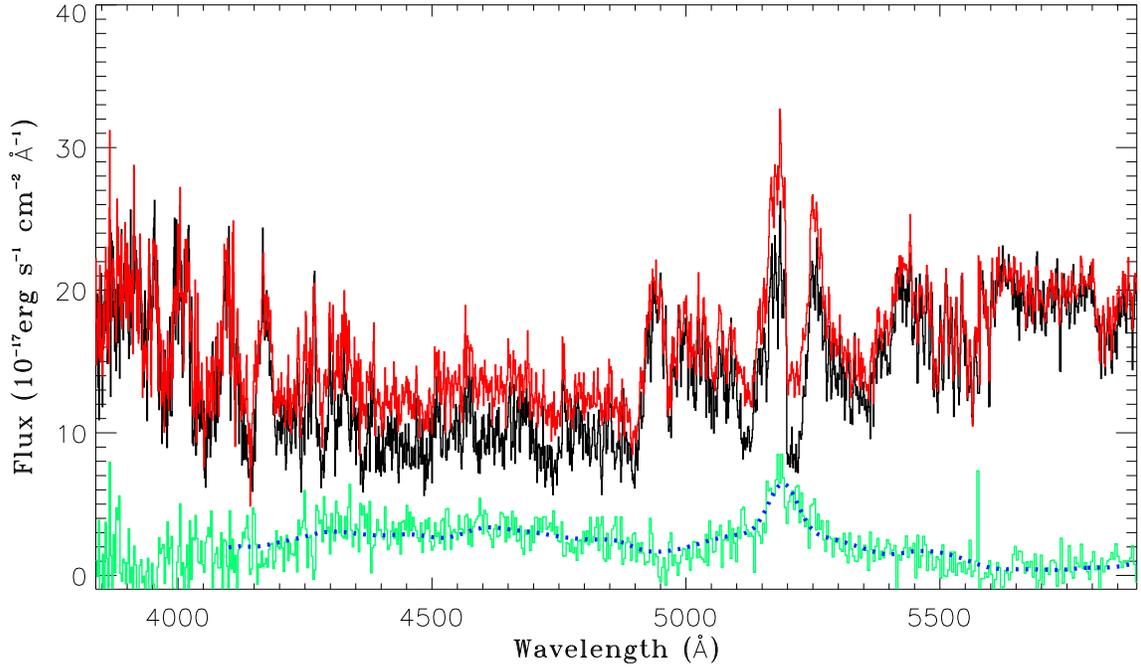}
\caption{Following the method in Hall \textit{et al.} (2013), we find that the fluxes of the scaled BOSS spectrum (red line) are more than $1\sigma$ larger than the fluxes of SDSS-I/II spectrum (black line) at the \ion{Mg}{2} and UV \ion{Fe}{2} troughs. The residual (green line) can be modeled using the emission of \ion{Mg}{2} and UV \ion{Fe}{2} $1130\ \mathrm{km\ s}^{-1}$ blueshifted to the quasar's rest frame with $\mathrm{FWHM}=4140\ \mathrm{km\ s}^{-1}$, as shown by the blue dotted line.\label{emis_var}}
\end{figure*}

Anyway, what we pay more attention to in this work is the absorption variability. In Hall \textit{et al.} (2013), the authors scaled the BOSS spectrum of J1125+0029 by a constant times a power-law to match the SDSS-I/II spectrum in absorption-free continuum regions near rest-frame 2100 and 2910 \AA. They suggested that the strengthened \ion{Mg}{2} emission alone is not sufficient to explain the change of \ion{Mg}{2} troughs. The absorption variability is still required, in which the blueshifted part of \ion{Mg}{2} absorption weakens more than the redshifted part. This seems not consistent with what we conclude from the Balmer and \ion{He}{1}* lines (see \S\ref{Measurement}). Thus, further investigation is necessary. In Figure \ref{emis_var} we plot the SDSS-I/II and scaled BOSS spectra following the method by Hall \textit{et al.} (2013). The fluxes of scaled BOSS spectrum are more than $1\sigma$ higher than those of SDSS spectrum at the bottom of \ion{Mg}{2} and UV \ion{Fe}{2} troughs. We find that if the assumed strengthened emission component is blueshifted to the quasar's rest frame by $1130\ \mathrm{km\ s}^{-1}$, the residual, plotted in green, can be well modeled if the \ion{Mg}{2} doublets and UV \ion{Fe}{2} emission template (constructed by Tsuzuki \textit{et al.} 2006) are convolved using a Guassian profile with $\mathrm{FWHM}=4140\ \mathrm{km\ s}^{-1}$. Therefore we think the strengthen of broad emission alone can explain the observed changes, and the absorption profiles for \ion{Mg}{2} and UV \ion{Fe}{2} could be assumed unchanged just like Balmer and \ion{He}{1}* lines.

\clearpage

\clearpage

\clearpage

\clearpage

\clearpage

\clearpage

\clearpage

\clearpage

\clearpage

\clearpage

\clearpage

\clearpage

\end{document}